\documentclass[12pt]{iopart} 
\usepackage{tikz}
\usepackage{graphicx}%
\usepackage{color} 
\usepackage{transparent}
\usepackage{psfrag}
\usepackage{dcolumn}
\usepackage{bm}
\usepackage{amssymb}
\usepackage{latexsym}
\usepackage{hyperref} 
\usepackage{booktabs}
\usepackage{latexsym, bbold} 
\begin{document}

\newcommand{\tikzcircle}[2][red,fill=red]{\tikz[baseline=-0.5ex]\draw[#1,radius=#2] (0,0) circle ;}%
\def\bea{\begin{eqnarray}}
\def\eea{\end{eqnarray}}
\def\beq{\begin{equation}}
\def\eeq{\end{equation}}
\def\f{\frac}
\def\k{\kappa}
\def\e{\epsilon}
\def\ve{\varepsilon}
\def\be{\beta}
\def\D{\Delta}
\def\h{\theta}
\def\t{\tau}
\def\a{\alpha}

\def\rpl{r_\parallel}
\def\rp{{\bf r}_\perp}

\def\cDa{{\cal D}[X]}
\def\cD{{\cal D}[x]}
\def\cL{{\cal L}}
\def\cLo{{\cal L}_0}
\def\cLa{{\cal L}_1}

\def\Re{{\rm Re}}
\def\sj{\sum_{j=1}^2}
\def\rk{\rho^{ (k) }}
\def\rek{\rho^{ (1) }}
\def\cek{C^{ (1) }}
\def\rz{\rho^{ (0) }}
\def\rt{\rho^{ (2) }}
\def\rtb{\bar \rho^{ (2) }}
\def\trk{\tilde\rho^{ (k) }}
\def\trek{\tilde\rho^{ (1) }}
\def\trz{\tilde\rho^{ (0) }}
\def\trt{\tilde\rho^{ (2) }}
\def\r{\rho}
\def\tD{\tilde {D}}

\def\s{\sigma}
\def\kb{k_B}
\def\bF{\bar{\cal F}}
\def\F{{\cal F}}
\def\la{\langle}
\def\ra{\rangle}
\def\nn{\nonumber}
\def\up{\uparrow}
\def\dn{\downarrow}
\def\S{\Sigma}
\def\dg{\dagger}
\def\d{\delta}
\def\p{\partial}
\def\l{\lambda}
\def\L{\Lambda}
\def\G{\Gamma}
\def\o{\Omega}
\def\w{\omega}
\def\g{\gamma}

\def\bv{ {\bf b}}
\def\uv{ {\hat {\bm{u}}}}
\def\rv{ {\bf r}}
\def\vv{ {\bf v}}

\def\jv{ {\bf j}}
\def\jr{ {\bf j}_r}
\def\jd{ {\bf j}_d}
\def\jdd{ { j}_d}
\def\noi{\noindent}
\def\a{\alpha}
\def\d{\delta}
\def\p{\partial} 

\def\la{\langle}
\def\ra{\rangle}
\def\e{\epsilon}
\def\n{\eta}
\def\g{\gamma}
\def\break#1{\pagebreak \vspace*{#1}}
\def\hf{\frac{1}{2}}

\title{Active Brownian particle in harmonic trap: exact computation of moments, and re-entrant transition }

\author{Debasish Chaudhuri}
\address{Institute of Physics, Sachivalaya Marg, Bhubaneswar 751005, India}
\address{Homi Bhaba National Institute, Anushaktigar, Mumbai 400094, India}
\ead{debc@iopb.res.in}

\author{Abhishek Dhar}
\address{International Centre for Theoretical Sciences, Tata Institute of Fundamental Research, Bengaluru 560089, India}
\ead{abhishek.dhar@icts.res.in}
\date{\today}

\begin{abstract}
We consider an active Brownian particle in a $d$-dimensional harmonic trap, in the presence of translational diffusion. While the Fokker-Planck  equation can not  in general be solved to obtain a closed form solution of the joint distribution of positions and orientations, as we show, it can be utilized to evaluate the exact time dependence of all  moments, using a Laplace transform approach. We present explicit calculation of several such moments at arbitrary times and their evolution to the steady state. In particular we compute the kurtosis of the displacement, a quantity  which clearly shows the difference of the active steady state properties from the equilibrium Gaussian form. We find that it increases with activity to asymptotic saturation, but varies non-monotonically with the trap-stiffness, thereby capturing  a recently observed  active- to- passive re-entrant behavior. 
\end{abstract}

\maketitle
\section{Introduction} 

Active matter describes systems of  self-propelled particles~\cite{Vicsek2012, Marchetti2013, Bechinger2016}. 
The associated energy pump and dissipation at the smallest scale maintains the system out of equilibrium, breaking the detailed balance condition and the equilibrium fluctuation-dissipation relation. Natural examples of active system range from  molecular motors, cytoskeleton, bacteria to bird, fish and animals. Many situations of biological interest involves confinement, e.g., chromosomes inside the cell nucleus~\cite{Alberts2007}.  

Drawing inspiration from the natural examples, a large number of artificial micro- or nano- swimmers, have been fabricated~\cite{Bechinger2016}.  The active colloids self-propel in their instantaneous heading directions, through auto-catalytic drive utilizing ambient optical, thermal, electrical, or chemical energy. 
Such a motion can be described by the active Brownian particle (ABP) model, where each particle self-propel with a constant speed in the heading direction, which undergoes stochastic 
reorientation. 
Two other related models, the run and tumble particles~\cite{Cates2013}, and the active Ornstein-Uhlenbeck process~\cite{Fodor2016,Das2018} show similar dynamics at long time scales. 
Active particles in confinement show qualitatively distinct properties with respect to passive Brownian particles. While, a dilute gas of passive particles distribute homogeneously within a confinement, that of active particles aggregate near the boundaries and corners~\cite{Wensink2008, Elgeti2009, Li2009b,Tailleur2009, Kaiser2012, Elgeti2013, Fily2014, Hennes2014, Solon2015c, Elgeti2016, Takatori2016, Li2017, Razin2017, Dauchot2019, Pototsky2012}.

In spite of the immense progress in the understanding of the collective properties of active matter, the non-equilibrium properties of non-interacting active particles is not yet completely understood. Even single self-propelled  particles can display rich and counterintuitive physical properties~\cite{Pototsky2012, Sevilla2014, Kurzthaler2016, Kurzthaler2018, maes2018, Basu2018, Basu2019, Shee2020, Majumdar2020, Basu2020}.   
Recent experiments~\cite{Takatori2016, Dauchot2019, Maggi2014, Li2009b},  and theoretical studies~\cite{Das2018, Basu2019, Malakar2018,  Dhar2019, Malakar2020, Duzgun2018,Wagner2017, Elgeti2015} of trapped active particles
have revealed hitherto unseen properties. These include a cross-over in the steady-state distribution of particle positions from a {\em passive} equilibrium-like Gaussian form with a peak at the trap-centre, to a strongly {\em active} non-Gaussian distribution with off-center peaks, as a function of trap stiffness and active velocity~\cite{Takatori2016, Basu2019, Malakar2020,Tailleur2009,Solon2015c}.

In this paper we reconsider the problem of ABPs in a harmonic trap. We develop an exact analytical method to calculate all time-dependent moments of the relevant dynamical variables, the active orientation and positional displacement of the ABP, in arbitrary $d$-dimensions.  This is the first main contribution of the paper. 
Remarkably,  the Fokker-Planck equation corresponding to the free ABP in the absence of translational noise was first studied as early as in 1952~\cite{Hermans1952, Daniels1952}, in the context of the worm-like-chain model of polymers.
Following Ref.~\cite{Hermans1952} we develop the Laplace transform method to calculate the exact moments
of the dynamical variables describing the ABP, in the presence of both the translational diffusion and a harmonic trap.
%
{
The presence of the trapping potential ensures that the system will eventually reach a steady state.}
We show the full evolution to steady state following the first few moments of the displacement vector. 
The kurtosis of the position vector quantifies its deviation from 
a {\em passive} equilibrium-like Gaussian distribution peaked at the trap-center. 
With increasing active velocity, the amplitude of kurtosis grows from zero to saturate monotonically. In contrast, with increasing trap stiffness, it shows a non-monotonic variation with the maximal amplitude appearing at an intermediate stiffness. We use the kurtosis to obtain a {\em phase diagram} identifying the {\em passive} and {\em active} phases in arbitrary $d$-dimensions, in terms of the amount of deviation from Gaussian statistics,  in the activity-trapping potential plane.
Note that there is no true phase transition in this single-particle system,  this is really a crossover between active and passive behavior. 
The phase diagram shows a re-entrant crossover, 
from a {\em passive} to {\em active}  to {\em passive} phase, 
with increasing trap stiffness, similar to what has  recently been reported for an ABP in a two-dimensional trap~\cite{Malakar2020}.  
Obtaining this phase diagram in $d$-dimensions analytically is the second main contribution of the current paper. 
In the context of the re-entrant behavior, we show that the presence of translational diffusion is crucial. 
In its absence the re-entrance disappears, and the amplitude of kurtosis grows monotonically to saturate with increasing trap stiffness. 

The plan of the paper is as follows.  
In Sec.~\ref{sec:model}, we present the model of ABP, and outline how to calculate any arbitrary moment of the relevant dynamical variables. In  Sec.~\ref{sec:moments}, we demonstrate detailed calculations of some of these moments, and analyze their time evolution. In Sec.~\ref{sec:kurtosis}, we describe the calculation of kurtosis, and present the phase diagram. Finally, in Sec.~\ref{sec:summary} we conclude, summarizing our main results.

\section{The Langevin and Fokker-Planck equations}
\label{sec:model}
{
Let us consider an active Brownian particles (ABP) in $d$ dimensions. It is  described by its position $\rv=(x_1,x_2,\ldots,x_d)$ and orientation $\uv=(u_1,u_2,\ldots,u_d)$, where $\uv.\uv=1$.  The orientation vector $\uv$ performs Brownian motion on the surface of a  $d$-dimensional hypersphere and drives the motion of the active particle. In the presence of a translational Brownian noise and an external potential,  $U(\rv)$, the particle's motion is described by the following stochastic equations (in Ito convention, see ~\cite{Ito1975,VandenBerg1985b,mijatovic2020}):
\bea
d x_i &=& \left[ v_0 u_i - \mu \partial_{x_i} U \right] dt + d{B_i^{\rm t}}(t) \label{eq:eom1}
\\
d u_i &=&  (\delta_{ij} -u_i u_j)d B_j^{\rm r} -{(d-1){D}_r} u_i dt,  \label{eq:eom2}
 \eea
where the Gaussian noise terms $ d{\bm B}^{\rm t}$ and $ d{\bm B}^{\rm r}$  have mean zero and variances  $\la d{B_i^{\rm t}}  d{B_j^{\rm t}} \ra = 2 D \delta_{ij} dt$, $\la d{B_i^{\rm r}}  d{B_j^{\rm r}} \ra =  2 {D}_r \d_{ij} dt$. Alternatively, we can write \Eref{eq:eom2} in the Stratonovich form $d u_i =  (\delta_{ij} -u_i u_j) \circ d B_j^{\rm r}$.  The form of \Eref{eq:eom2} ensures the normalization $\uv^2=1$ at all times.

Noting that the stochastic equations, Eqs.~(\ref{eq:eom1},\ref{eq:eom2}), correspond to the orientational vector $\uv$ performing Brownian motion on the surface of a unit $d$-dimensional hypersphere while the position vector $\rv$ evolves via standard drift and diffusion terms, we can write the corresponding  Fokker-Planck equation  for the probability distribution $P(\rv, \uv, t)$: 
\bea
\p_t P(\rv, \uv, t) = D \nabla^2 P  -  \nabla  \cdot [  v_0 \uv - \mu \nabla  U]   P   + D_r \nabla_u^2 P, 
\label{eq:FP1} 
\eea
where $ \nabla = (\partial_{x_1},\partial_{x_2},\ldots,\partial_{x_d}) $ denotes the $d$-dimensional gradient operator, and $\nabla_u^2$ denotes the spherical Laplacian defined on the ($d-1$) dimensional orientation space. We note that the spherical Laplacian can be expressed in terms of the cartesian coordinates ${\bf y}$, defined through $u_i=y_i/y$ where $y=|{\bf y}|$, as $\nabla^2_u= y^2\sum_{i=1}^d \partial^2_{y_i} - [y^2\partial^2_y + (d-1) y \partial_y]$.
}
In the limit of $D=0$ and $U(\rv) = 0$, the above equation can be interpreted as that
describing the probability distribution of the end-to-end separation $\rv$ of a worm-like-chain with bending rigidity $\k \sim 1/D_r$, interpreting the polymer contour length $L \sim t$~\cite{Hermans1952, Dhar2002, Chaudhuri2007, Kurzthaler2018a, Shee2020}.    
For a harmonic trap, $U (\rv) = (1/2) k \rv^2$, the \Eref{eq:FP1} simplifies to 
\bea
\p_t P(\rv, \uv, t) = D \nabla^2 P + D_r \nabla_u^2 P - v_0\, \uv\cdot  \nabla  P
+ \mu k \rv \cdot  \nabla  P + \mu k d\, P.  \nn
\eea
Using the Laplace transform $\tilde P(\rv, \uv, s) = \int_0^\infty dt e^{-s t} P(\rv, \uv, t) $, this Fokker-Planck  equation can be recast into the form,
\bea
\fl
-P(\rv, \uv, 0) + (s-\mu k d) \tilde P(\rv, \uv, s) = D \nabla^2 \tilde P + D_r \nabla_u^2 \tilde P 
- v_0\, \uv\cdot \nabla \tilde P + \mu k \rv \cdot \nabla \tilde P.  \nn
\eea
 Defining the mean of an observable in the Laplace space $\la \psi \ra_s = \int d\rv \, d\uv\, \psi(\rv, \uv ) \tilde P(\rv, \uv, s)$, and multiplying the above equation by $\psi(\rv, \uv)$ and integrating over all possible $(\rv, \uv)$  we find,
\bea
\fl
-\la \psi \ra(0) + (s-\mu k d) \la \psi \ra_s = D \la \nabla^2 \psi \ra_s + D_r \la \nabla_u^2 \psi \ra_s  
 + v_0\, \la \uv\cdot \nabla \psi \ra_s - \mu k \la  \nabla \cdot (\rv \psi) \ra_s.
\label{moment2}
\eea
where the initial condition sets $\la \psi \ra(0) = \int d\rv \, d\uv\, \psi(\rv, \uv) P(\rv, \uv, 0)$. 
{
We consider the initial position of the particle to be at $\rv_0$, shifted from the center of the trap, and its initial orientation of activity along $\uv_0$ such that  $P(\rv, \uv, 0) =  \d(\rv - \rv_0) \d(\uv - \uv_0)$. In the following section we show how  \Eref{moment2} can be used to determine various  moments of interest.}

\section{Calculation of moments}
\label{sec:moments}
In this section, we present detailed derivation of some of the relevant moments, show their time evolution, and analyze
their steady state properties.

\subsection{Orientational correlation}
 The confinement is not expected to change the orientational relaxation of activity in a spherically symmetric particle. This can be illustrated by directly computing the two time correlation function of the active orientation. 
We use $\psi(\rv, \uv) = \uv \cdot \uv_0$ in \Eref{moment2}, to find $\la \psi\ra(0) = 1$, $\la \nabla^2 \psi \ra_s =0$, $\la \nabla_u^2 \psi \ra_s = - (d-1) \uv \cdot \uv_0$, and $\la \uv \cdot \nabla \psi \ra_s = 0 $, and $ \la \nabla  \cdot  (\rv \psi) \ra_s = d \la \uv \cdot \uv_0 \ra_s$. This leads to the relation
\bea
\la \uv \cdot \uv_0 \ra_s = \f{1}{s +(d-1)D_r},
\eea 
which, after performing an inverse Laplace transform leads to
\bea
\la \uv \cdot \uv_0 \ra (t) = e^{-  (d-1)D_r\, t}.
\eea
 As expected, it gives an exponential decay of the two-time orientational correlation in $d$-dimensions with a correlation time $\t_c = [ (d-1) D_r ]^{-1}$, independent of the confinement. 

We use $\t_r = 1/D_r$ to set the unit of time in the problem. Along with the translational diffusion constant $D$, this gives the 
unit of length $\bar \ell = \sqrt{D/D_r}$. The dimensionless active velocity $\l = v_0/\sqrt{D D_r}$ and strength of the trap $\be = \mu k/D_r$ control the various properties of the ABP in harmonic trap. {
Note that the dimensionless activity $\l$ is equivalent to the P{\'e}clet number defined as $v_0/\bar\ell D_r$.}

\begin{figure}[!t]
\centering
\includegraphics[width=10cm]{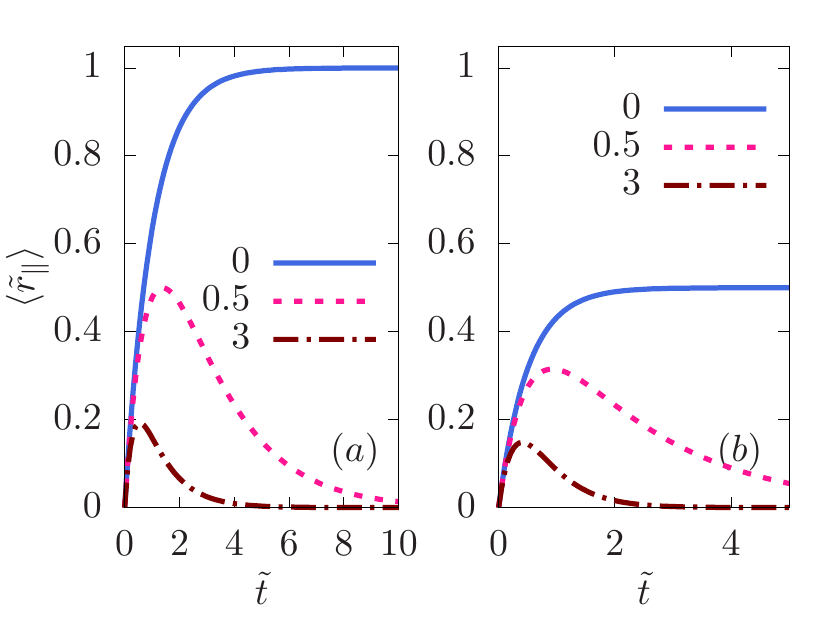}
\caption{(color online)
Dependence of displacement parallel to the initial orientation $\la \tilde \rpl \ra$ on time $\tilde t$, in $d=2$\,($a$) and $d=3$\,($b$) dimensions.
We use dimensionless activity  $\l=1$. 
The three curves in each of these figures correspond to 
dimensionless  strengths  of harmonic potentials $\be = 0,\, 0.5, \, 3$. The initial position ${\tilde\rv}_0$ was considered at the origin, coinciding the center of the trap.}
\label{fig:rpl}
\end{figure}

\subsection{Displacement vector}
Using $\psi = \rv$ in \Eref{moment2}, we get for the displacement vector 
$\la \rv \ra_s = \f{v_0 \la \uv \ra_s  {\color{blue} +\rv_0}}{s + \mu k}$.
The same \Eref{moment2}, with $\psi=\uv$ gives
$\la \uv \ra_s = \f{\uv_0}{s+(d-1)D_r}.$
Thus, together, we obtain
\bea
\la \rv \ra_s = \f{v_0 \uv_0}{ (s+\mu k)(s + (d-1)D_r )} { + \f{\rv_0}{(s+\mu k)} },
\eea
which under inverse Laplace transform leads to
\bea
\la \rv \ra\, (t) = \f{v_0 \hat u_0}{-\mu k + (d-1)D_r }\left(  e^{-\mu k t} - e^{-(d-1)D_r t} \right)
 + \rv_0 e^{-\mu k t}  . 
\eea
The displacement components parallel and perpendicular to the initial orientation $\uv_0$ are defined as $\rv_\parallel = \rpl \uv_0$ with $\rpl = \rv \cdot \uv_0$, and $\rv_\perp = \rv - \rv_\parallel$, respectively. By symmetry, 
$\la \rp \ra = (r_0)_\perp \exp(-\mu k t)$.
In terms of dimensionless form $\la {\tilde r}_\parallel \ra = \la \rpl \ra/\bar \ell$ we find 
\bea
\la {\tilde r}_\parallel \ra = \f{\l }{-\be + (d-1)} \left(e^{-\be \tilde t} - e^{-(d-1)\tilde t}\right) 
 + ({\tilde r}_0)_\parallel e^{-\be \tilde t} ,
\label{eq:rpl}
\eea
where, $\tilde t = t \, D_r$. At short time, $\tilde t \ll 1$, the displacement grows linearly with time $\la  {\tilde r}_\parallel \ra = \l \tilde t$. However, in the limit $\tilde t \gg 1$, the harmonic trap ensure that the steady state displacement
$\la {\tilde r}_\parallel \ra = 0$. The time-scale over which the parallel component of the displacement vector vanishes is given by $\be^{-1}$, i.e., stronger the trap is the faster is the vanishing. This 
dependence for both $d=2$ and $3$ are shown in Fig.~\ref{fig:rpl}.

\subsection{Second moment of displacement}
Considering the displacement squared as the dynamical variable of interest,  $\psi = \rv^2$, and using  in \Eref{moment2} the related facts that  
$\la \psi \ra(0) = \rv_0^2$,
$\la \nabla_u^2 \psi \ra_s = 0$, $\la \nabla^2 \psi \ra_s = 2d/s$, $\la \uv \cdot \nabla \psi \ra_s = 2 \la \uv \cdot \rv \ra_s$, and $\la   \nabla \cdot  (\rv \psi )\ra_s = (d + 2) \la \rv^2 \ra_s$, one finds 
\bea
\la \rv^2 \ra_s = \f{1}{s +2 \mu k} \left[  \f{2 d D}{s}  + 2 v_0 \la \uv \cdot \rv\ra_s  + \rv_0^2 \right].
\label{rv2s}
\eea
Further, using $\psi= \uv \cdot \rv$, and the results $\nabla^2 (\uv \cdot \rv) = 0$, $\nabla_u^2 (\uv \cdot \rv) = -(d-1) (\uv \cdot \rv)$, $\la \uv \cdot \nabla(\uv \cdot \rv) \ra_s = \la \uv^2 \ra_s = \la 1 \ra_s = 1/s $, and $\nabla \cdot [\rv  (\uv \cdot \rv)] = (d+1) (\uv \cdot \rv)$ we obtain,
\bea
\la \uv \cdot \rv \ra_s = \f{1}{s + \mu k +(d-1) D_r}~\left( \f{v_0}{s} + \uv_0 \cdot \rv_0 \right).
\label{udotr2}
\eea
Using \Eref{udotr2} in \Eref{rv2s} one obtains,  
\bea
\fl
\la \rv^2 \ra_s =  \f{1}{s +2 \mu k} \left[  \f{2dD}{s}  +\f{2 v_0^2}{s \{ s + \mu k +(d-1) D_r\}} 
{ + \f{2 v_0 \, \,\uv_0 \cdot \rv_0}{s + \mu k +(d-1) D_r} + \rv_0^2} \right] .
\label{rvsq2}
\eea
The inverse Laplace transform of \Eref{udotr2} provides the time-dependence of the equal time cross-correlation between the active orientation and the displacement vector of the ABP 
\bea
\fl
\la \uv \cdot \rv \ra (t) = \f{v_0}{\mu k+ (d-1) D_r } \left( 1 - e^{-[ \mu k + (d-1) D_r ] t} \right) 
{ +  (\uv_0 \cdot \rv_0) \, e^{-[\mu k + (d-1) D_r] t}},
\label{eq:ru1}
\eea 
which depends on the strength of the trapping potential. 
This correlation in the steady state can be directly obtained from the above relation, or using the final value theorem, 
\bea
\lim_{t \to \infty} \la \uv \cdot \rv \ra (t) = \lim_{s \to 0_+}  s\la \uv \cdot \rv \ra_s = \f{v_0}{\mu k +(d-1) D_r},
\label{eq_ur}
\eea
{which shows that orientations and positions are correlated in the steady state.} Performing the inverse Laplace transform of \Eref{rvsq2} one gets,
\bea
\fl
\la \rv^2(t) \ra = \f{d\, D}{\mu k} \left(1 - e^{-2\mu k t} \right) +  \f{v_0^2}{\mu k[ (d-1) D_r + \mu k ] } 
 - \f{2 v_0^2 e^{-\mu k t}}{(d-1) D_r - \mu k } \left[ \f{e^{-\mu k t} }{2 \mu k}  - \f{e^{- (d-1) D_r t}}{ (d-1) D_r + \mu k }\right]\nn \\
 { + \f{2 v_0 \, \uv_0\cdot \rv_0}{\mu k- (d-1)D_r} \left( e^{-[ \mu k + (d-1) D_r]t} - e^{- 2 \mu k t} \right) + \rv_0^2 e^{- 2 \mu k t} }
 \label{eq:r2full}
\eea
The dimensionless form, $\la {\tilde \rv}^2 \ra = \la \rv^2(t) \ra/ \bar \ell^2$,  can be expressed as
\bea
\la {\tilde \rv}^2(\tilde t) \ra &=& \f{d}{\be} \left(1 - e^{-2\be \tilde t} \right) +  \f{\l^2}{\be [ d-1+ \be ] }
 - \f{2 \l^2 e^{-\be \tilde t}}{d-1 - \be } \left[ \f{e^{-\be \tilde t} }{2 \be}  - \f{e^{- (d-1) \tilde t}}{ d-1+ \be }\right] \nn\\
  &+& { \f{2 \l \, \uv_0\cdot {\tilde \rv_0}}{\be- (d-1)} \left( e^{-[ \be + (d-1)]\tilde t} - e^{- 2 \be \tilde t} \right) + {\tilde \rv_0}^2 e^{- 2 \be \tilde t} }
\label{eq:rv2}
\eea
\begin{figure}[!t]
\centering
\includegraphics[width=10cm]{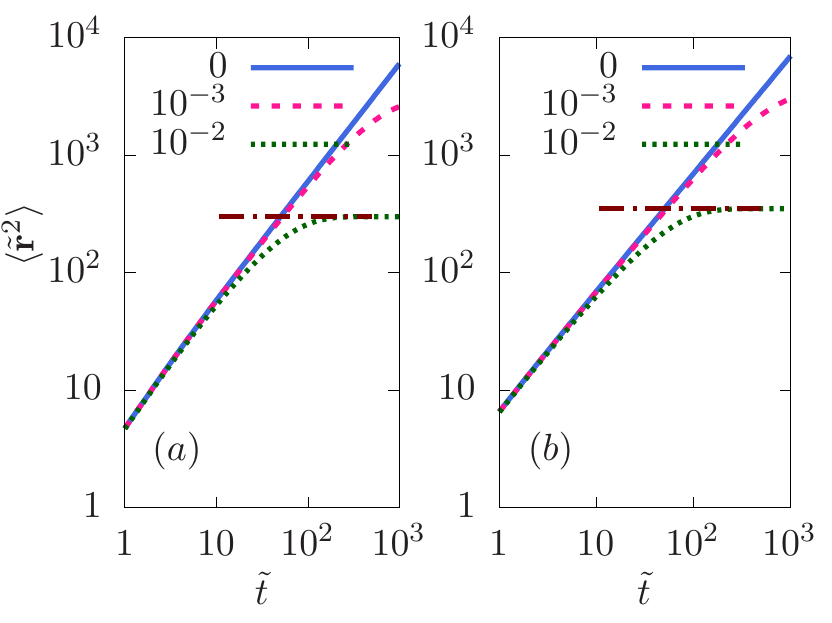}
\caption{(color online)
Evolution of the dimensionless form of the second moment of displacement $\la {\tilde \rv}^2(t) \ra$, in dimensions $d=2$~($a$) and $d=3$~($b$) at dimensionless  strengths  of harmonic potentials $\be = 0,\, 10^{-3}, \, 10^{-2}$ indicated by the three lines in the figures. 
The strength of activity $\l=1$ is used. 
The dash-dotted lines in ($a$) and ($b$) denote the asymptotic steady state 
$\lim_{\tilde t \to \infty} \la {\tilde \rv}^2 \ra$ at $\be=10^{-2}$, as described by Eq.~\ref{eq:rv2steady}. The initial position ${\tilde \rv}_0$ was considered at the center of the trap.
}
\label{fig:rv2}
\end{figure}
The second moment of displacement evolves to the steady state value  
\bea
{\tilde \rv_{st}}^2 = \lim_{\tilde t \to \infty} \la {\tilde \rv}^2 \ra = \f{1}{\bar \ell^2} \lim_{s \to 0_+}  s\la { \rv}^2 \ra_s = 
\f{d}{\be} + \f{\l^2}{\be( d-1 + \be)}. 
\label{eq:rv2steady}
\eea
%
It is easy to check that in the limit of vanishing trap-stiffness 
\Eref{eq:rv2} reduces to 
\bea
\fl
\la {\tilde \rv}^2(\tilde t) \ra = 2d \left( 1 + \f{\l^2}{d(d-1)}\right) \tilde t  - \f{2 \l^2}{(d-1)^2} \left( 1  - e^{- (d-1) \tilde t} \right) 
{ +  \f{2\l\, \uv_0 \cdot {\tilde \rv_0}}{d-1} \left(1 - e^{-(d-1)\tilde t} \right) + {\tilde \rv_0}^2.} \nn\\
\label{eq:rv2free}
\eea
{
Setting $\rv_0$ as the origin, Eq.~(\ref{eq:r2full}) then reduces to the known result for free ABP in $d$-dimensions~\cite{Elgeti2015a, Shee2020}, 
\bea
\la \rv^2 \ra = 2 d D_{\rm eff} t - \f{2 v_0^2}{(d-1)^2 D_r^2} \left( 1-e^{-(d-1)D_r t}\right),
\eea
with { the effective diffusion coefficient} 
\bea 
D_{\rm eff} = D + \f{v_0^2}{d(d-1)D_r}, 
\label{eq_Deff}
\eea
{ describing the long time diffusive behavior.}
This is in agreement with the $d=2$ dimensional result in Ref.~\cite{Howse2007}. In the limiting case of $D=0$, the expression of the second moment of displacement reduces to the well-known result of the end-to-end separation in the worm-like-chain model~\cite{Dhar2002,Hermans1952},
\bea
\la \rv^2 \ra = \f{4\k l}{d-1} - \f{8 \k^2(1-e^{-\f{(d-1)l}{2\k}})}{(d-1)^2},
\eea
where we identify the bending rigidity $\k=v_0/2D_r$ and chain length $l=v_0 t$.}

The evolutions described by \eref{eq:rv2} and \Eref{eq:rv2free}
are shown in Fig.~\ref{fig:rv2}. In the absence of the harmonic potential, $\be = 0$, the moment $\la \rv^2 \ra$ show two crossovers,
from diffusive $\sim t$ behavior at shortest times to ballistic $\sim t^2$ behavior that again crossover to long time diffusion $\sim t$~\cite{Shee2020}. 
The trapping potential brings back the displacement moment towards the steady state value described by Eq.~(\ref{eq:rv2steady}). In Fig.~\ref{fig:rv2}, we show evolution of the scaled mean squared position $\la {\tilde \rv}^2 \ra $ for different trapping potential in 2d and 3d starting from the initial position $\rv_0$ at the center of the trap.  

\begin{figure}[!t]
\centering
\includegraphics[width=16cm]{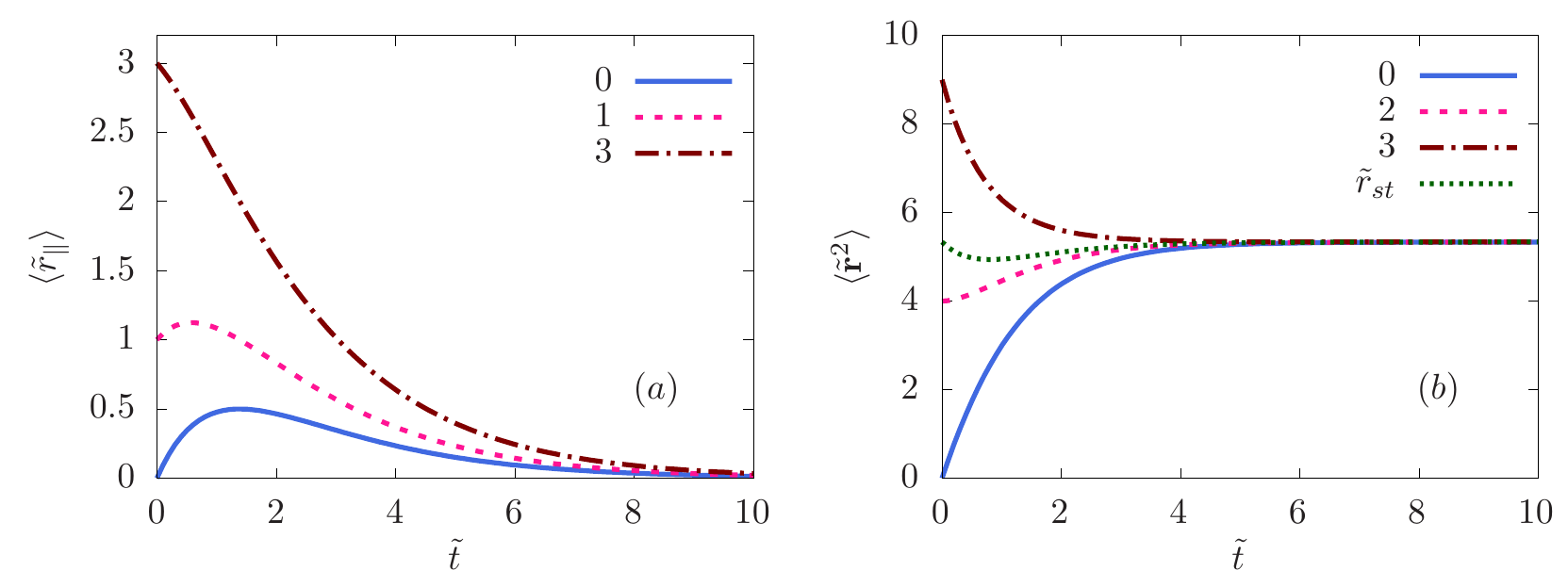}
\caption{{(color online) ($a$)~Dependence of $\la \tilde \rpl \ra(t)$ on initial positions $ ({\tilde r}_0)_\parallel= \uv_0 \cdot \rv_0/\bar \ell$, the values of which are denoted in the figure legend. ($b$)~Dependence of $\la {\tilde \rv}^2 \ra(t)$, averaged over all possible initial orientations $\uv_0$, on separation $\tilde r_0$ of the initial position from the center of the trap. 
The plots are shown at $d=2$ dimensions, using parameter values $\l=1,\,\be=0.5$. 
Even with a choice of initial separation ${\tilde r}_0 = \tilde r_{st}$ having the steady state value $ \tilde r_{st}=\sqrt{\tilde\rv_{st}^2} =2.31$~(Eq.(\ref{eq:rv2steady})\,), the evolution of $\la {\tilde \rv}^2 \ra(t)$ shows transient deviations before returning to the steady state value~(green dotted line).}  
}
\label{fig:rv_rv2}
\end{figure}

{
Although the steady state in the presence of the harmonic trap  is independent of  initial conditions,  the time-evolution of individual trajectories depend on the initial position.  In Fig.~\ref{fig:rv_rv2} we show the evolution starting from  various initial positions. In Fig.~(\ref{fig:rv_rv2}a), we plot the displacements parallel to the initial orientation of activity while in Fig.~(\ref{fig:rv_rv2}b), the second moment of displacement is plotted after taking an average over all possible initial orientations of activity.  It is interesting to note that even if the initial value of position $\tilde\rv_0^2$ is chosen to have the steady state  value $\tilde \rv_{st}^2$, the moment shows transient deviations before evolving back to the steady state value. { This transient deviation can be understood from Eq.(\ref{eq:rv2}). The orientational averaging renders $\la \uv_0 \cdot \tilde \rv_0 \ra=0$. The contribution from initial position $\tilde r_0^2$ decays in time scale $(2\beta)^{-1}$, and in that same time-scale the first term in the expression of $\la \tilde r^2 (\tilde t) \ra$ saturates. The third term contributes to the transient deviation, as well, vanishing at late times. }
}

\subsection{Fourth moment of displacement}
Again to calculate  $\la \rv^4 \ra (t)$, we consider $\psi = \rv^4$ and use \Eref{moment2}. Note 
{$\la \rv^4 \ra(0) = \rv_0^4$},
$\la \nabla_u^2 \rv^4 \ra_s = 0$. It is straightforward to show that 
$\la \nabla^2 \rv^4 \ra_s = 4(d+2) \la \rv^2 \ra_s$. Further, $\la \uv \cdot \nabla \rv^4 \ra_s = 4 \la \uv \cdot \rv^2 \rv \ra_s$, and 
$\la \nabla \cdot (\rv \psi) \ra_s = (d+4) \la \rv^4 \ra_s$. Thus one gets
\bea
\la \rv^4 \ra_s =  \f{4 (d+2) D}{s + 4\mu k} \la \rv^2 \ra_s + \f{4 v_0}{s + 4 \mu k} \la (\uv \cdot \rv) \rv^2 \ra_s 
{+ \f{\rv_0^4}{s + 4\mu k}}. 
\label{step1}
\eea
Now we proceed to calculate $\la \psi \ra_s$ with $\psi =  (\uv \cdot \rv) \rv^2$. We find, 
{ $\la \psi \ra_0 = (\uv_0 \cdot \rv_0) \rv_0^2$,}
$\la \nabla^2 \psi \ra_s = 2(d+2) \la \uv \cdot \rv \ra_s$. 
 Using $\nabla_u^2 u_i = -(d-1) u_i$ one finds $\la \nabla_u^2 \psi \ra_s = -(d-1) \la (\uv \cdot \rv) \rv^2\ra_s$. 
The self propulsion term $\la \uv \cdot \nabla \psi \ra_s = \la \rv^2\ra_s + 2 \la (\uv \cdot \rv)^2 \ra_s$. The final term $\la \nabla \cdot(\rv \psi)\ra_s = (3+d ) \la  (\uv \cdot \rv) \rv^2 \ra_s$. 
This leads to the relation
\bea
\fl
\la (\uv \cdot \rv) \rv^2 \ra_s = \f{2(d+2)D\, \la \uv \cdot \rv \ra_s}{s+3 \mu k + (d-1)D_r}   
 + \f{v_0 \la \rv^2 \ra_s + 2 v_0 \la (\uv \cdot \rv)^2 \ra_s}{s+3 \mu k + (d-1)D_r}
 { + \f{(\uv_0 \cdot \rv_0) \rv_0^2}{s+3 \mu k + (d-1)D_r}}
\label{step2}
\eea
Note that $\la \uv \cdot \rv \ra_s$, and $\la \rv^2 \ra_s$ have already been calculated in Eq.~(\ref{udotr2}), and Eq.~(\ref{rvsq2}), respectively. 
We are left to calculate  $\la \psi \ra_s$ with $\psi =  (\uv \cdot \rv)^2$, which gives 
{ $\la \psi \ra(0) = (\uv_0 \cdot \rv_0)^2$,}
$\nabla^2 \psi = 2 \uv^2 =2$, and thus $\la \nabla^2 \psi \ra_s = 2/s$. 
Given $\nabla_u^2 u_m u_p = -2 d u_m u_p +2\d_{mp}$ we find $\la \nabla_u^2 \psi \ra_s = 2 \la \rv^2 \ra_s - 2 d \la (\uv \cdot \rv)^2 \ra_s$. The self propulsion term
$\la \uv \cdot \nabla \psi \ra_s = 2 \la \uv \cdot \rv \ra_s$. The final term associated with the trapping potential $\la \nabla \cdot (\rv \psi) \ra_s = (d+2) \la (\uv \cdot \rv)^2 \ra_s$.
Thus we obtain the relationship,
\bea
\fl
(s + 2 \mu k + 2 d D_r)\la (\uv \cdot \rv)^2 \ra_s = \f{2D}{s} + 2 D_r \la \rv^2 \ra_s  
 + 2 v_0 \la \uv \cdot \rv \ra_s
{ + (\uv_0 \cdot \rv_0)^2} .
\label{step3}
\eea 
Using Eqs.~(\ref{step1}), (\ref{step2}), and (\ref{step3}) one finally obtains the expression for 
\bea
\fl
\la \rv^4 \ra_s = \frac{8 d (d+2) D^2}{s (2 \mu k+s) (4 \mu k+s)}
+\frac{8 (d+2) D v_0^2}{s (2 \mu k+s) (4 \mu k+s) ((d-1) D_r+\mu k+s)} \nn\\
+\frac{8 (d+2) D v_0^2 }{s (4 \mu k+s) ((d-1) D_r+\mu k+s) ((d-1) D_r+3 \mu k+s)}\nn\\
+\frac{8 d D v_0^2}{s (2 \mu k+s) (4 \mu k+s) ((d-1)  D_r+3 \mu k+s)}\nn\\
+\frac{16 D v_0^2}{s (4 \mu k+s) (2 d D_r+2 \mu k+s) ((d-1) D_r+3 \mu k+s)}\nn\\
+\frac{32 d D D_r v_0^2}{s (2 \mu k+s) (4 \mu k+s) (2 d D_r+2 \mu k+s) ((d-1) D_r+3 \mu k+s)}\nn\\
+\frac{8 v_0^4}{s (2 \mu k+s) (4 \mu k+s) ((d-1) D_r+\mu k+s) ((d-1) D_r+3 \mu k+s)}\nn\\
  +\frac{16 v_0^4}{s (4 \mu k+s) ((d-1) D_r+\mu k+s) (2 d D_r+2 \mu k+s) ((d-1) D_r+3 \mu k+s)}\nn\\
+\frac{32 D_r v_0^4}{s (2 \mu k+s) (4 \mu k+s) ((d-1) D_r+\mu k+s) (2 d D_r+2 \mu k+s) ((d-1) D_r+3 \mu k+s)} \nn\\
{ + \f{\rv_0^4}{s+4\mu k} + \f{8(d+2)D\,\, v_0 \uv_0\cdot\rv_0}{(s+4\mu k)(s+\mu k + (d-1)D_r)} \left( \f{1}{s+2\mu k}+ \f{1}{s+3\mu k +(d-1)D_r}\right) } \nn\\
{+ \f{ {\rv_0}^2}{s+4\mu k} \left( \f{4(d+2)D}{s+2\mu k}       %
+ \f{4 v_0(\uv_0 \cdot \rv_0)}{s+3\mu k+(d-1)D_r} 
+\f{8 {v_0}^2}{(s+3\mu k +(d-1)D_r)(s+2\mu k + 2d D_r)} \right)} \nn\\
\eea
Performing inverse Laplace transform,  the dimensionless fourth moment $\la {\tilde \rv}^4 \ra =  \la \rv^4 \ra (t) /\bar \ell^4$ can be expressed as
\bea
\fl
\la {\tilde \rv}^4(\tilde t) \ra =  \frac{d (d+2)}{\be^2} + \frac{2(d+2) \l^2}{\be^2 (d-1+\be)} +\frac{(2+d+3 \be) \l^4}{\be^2  (d-1+\be) (d+\be) (d-1+3 \be)} \nn\\
+\frac{2 (d+2)  \left(d^2 -d (1+\be)+\l^2\right) \left(d (d-1+\be)+\l^2\right)}
{\be^2 d  \left( \be^2 - (d-1)^2  \right)} e^{-2 \be \tilde t}\nn\\
+\frac{e^{-4 \be \tilde t} } {\be^2 (\be-d ) ( 1 +\be -d ) (1+3 \be -d )}
\left[-d (d+2)  (d-1 -3   \be) (d-1-\be) (d -\be) \right.\nn\\
\left. -2 (d+2)   ( d-1 -3 \be) (d-\be) \l^2 -  (d+2 -3 \be) \l^4 \right] \nn\\
 - \frac{4 \l^2  \left((d+2) (d +1-\be) (d-1+3 \be)+\l^2 (d-7+3  \be)\right)}
     {\be (d-1-\be) (d +1-\be) (d-1+\be) (d-1+ 3 \be)} e^{-(d -1+3 \be) \tilde t } \nn\\
 +\frac{4 \l^2 \left( (d+2)  (d +1+\be)  (d-1 -3 \be) +\l^2 (d-7 -3 \be)\right)}
   {\be (d-1-3 \be) (d-1-\be) (d-1+\be) (d +1+\be)} e^{-( d-1 +\be) \tilde t } \nn\\
 +\frac{4 (d-1) \l^4 }{d (d -\be) (d +1-\be) (d +\be) (d + 1+\be)} e^{-2  (d +\be) \tilde t}
  { + {\tilde \rv}_0^4 e^{-4 \be \tilde t} }\nn\\
  {  + \f{4 (d+2)}{2 \be}\, {\tilde \rv_0}^2 \left( e^{-2\be\tilde t} - e^{-4 \be \tilde t}\right) 
  + \f{4 \l(\uv_0\cdot {\tilde \rv}_0) {\tilde \rv}_0^2}{\be - d+1} \left( e^{-(3\be+d-1)\tilde t} - e^{-4 \be \tilde t}\right)} \nn\\
   {  + 8 \l^2 {\tilde\rv}_0^2 \left[  \f{e^{-4 \be \tilde t}}{(-\be +d-1)(-2\be +2d)} + \f{e^{-(3\be + d -1)\tilde t} }{(\be -d+1)(-\be+d+1)} + \f{e^{-(2\be+2d)\tilde t}}{(2 \be -2 d)(\be-d-1)} \right] } \nn\\
   {  +  8(d+2) \l \uv_0 \cdot {\tilde\rv}_0 \left[ \f{e^{-4\be \tilde t}}{(-3\be + d -1)(-2\be)} 
   + \f{e^{-(\be + d -1)\tilde t}}{(3 \be -d+1)(\be - d +1) } + \f{e^{-2 \be \tilde t}}{(2\be)(-\be+d-1)} \right.}  \nn\\ 
   { \left. +\f{e^{-4 \be \tilde t}}{(-3\be +d-1)(-\be+d-1)} + \f{e^{-(\be+d-1)\tilde t}}{(3\be-d+1)(2\be)} + \f{e^{-(3\be + d-1)\tilde t}}{(\be-d+1)(-2\be)}   \right] }
\label{eq:rv4}
\eea
%
\begin{figure}[!t]
\centering
\includegraphics[width=10cm]{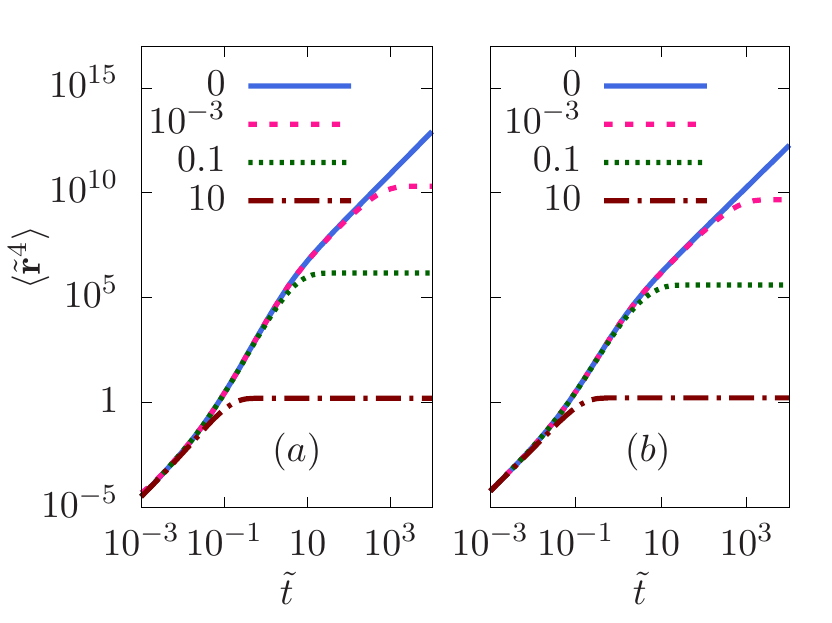}
\caption{(color online)
Evolution of the fourth moment of displacement $\la {\tilde \rv}^4(t)\ra$ in dimensions $d=2$~($a$) and $d=3$~($b$) at 
 harmonic potential strengths $\be = 0,\, 10^{-3}, \, 0.1,\, 10$ indicated by the four lines in the figures. In this figure $\rv_0$ is assumed at the origin, and we used the strength of activity $\l=10$. The initial position ${\tilde \rv}_0$ was considered at the center of the trap.
}
\label{fig:rv4}
\end{figure}
%
In the limit of vanishing trap stiffness, { using the initial position $\rv_0$ at the origin,} the  above relation leads to  
\bea
\fl
\la {\tilde \rv}^4(\tilde t) \ra = \frac{4 (d-1) \l^4 e^{-2 d \tilde t}}{d^3 (d+1)^2 } 
-\frac{8 \l^4 \left(d^2 +10 d +25 \right) e^{-(d-1) \tilde t}}{(d-1)^4 (d+1)^2}
+\frac{4 \l^4 \left(d^3 +23 d^2 -7 d +1\right)}{(d-1)^4 d^3}\nn\\
+\frac{8 \tilde t e^{- (d-1) \tilde t } \left(d^3 \l^2+2 d^2  \l^2-d \l^2+d \l^4-2  \l^2-7 \l^4\right)}{(d-1)^3 (d+1)}\nn\\
   +\frac{4 {\tilde t}^2 \left(d^5 -3 d^3 +2 d^3  \l^2+2 d^2 +2 d^2  \l^2-4 d  \l^2+d \l^4+2 \l^4\right)}{(d-1)^2 d }\nn\\
   -\frac{8 \tilde t \left(d^4 \l^2+d^3  \l^2-2 d^2  \l^2+d^2 \l^4+6 d \l^4-\l^4\right)}{(d-1)^3 d^2}
   \label{eq:rv4free}
 \eea
a result obtained before in Ref.~\cite{Shee2020}.

In the presence of the external trapping potential the fourth moment of displacement reaches a steady state value. It is relatively simple to obtain this expression using the 
final value theorem.  
The expression is given by, 
\bea
\fl
\lim_{t \to \infty} \la {\tilde \rv}^4 \ra= \f{1}{\bar \ell^4}\lim_{s \to 0_+} s \la \rv^4 \ra_s = 
\frac{d (d+2)}{\be^2} + \frac{2(d+2) \l^2}{\be^2 (d-1+\be)} +\frac{(2+d+3 \be) \l^4}{\be^2  (d-1+\be) (d+\be) (d-1+3 \be)}. \nn\\
\eea
The evolution of the fourth moment with time is shown in Fig.~\ref{fig:rv4} at different strengths of the trapping potential $\be$, using \Eref{eq:rv4} and \Eref{eq:rv4free}. 


\section{Deviation from Gaussian nature: re-entrance}
\label{sec:kurtosis}
\begin{figure}[!t]
\centering
\includegraphics[width=12cm]{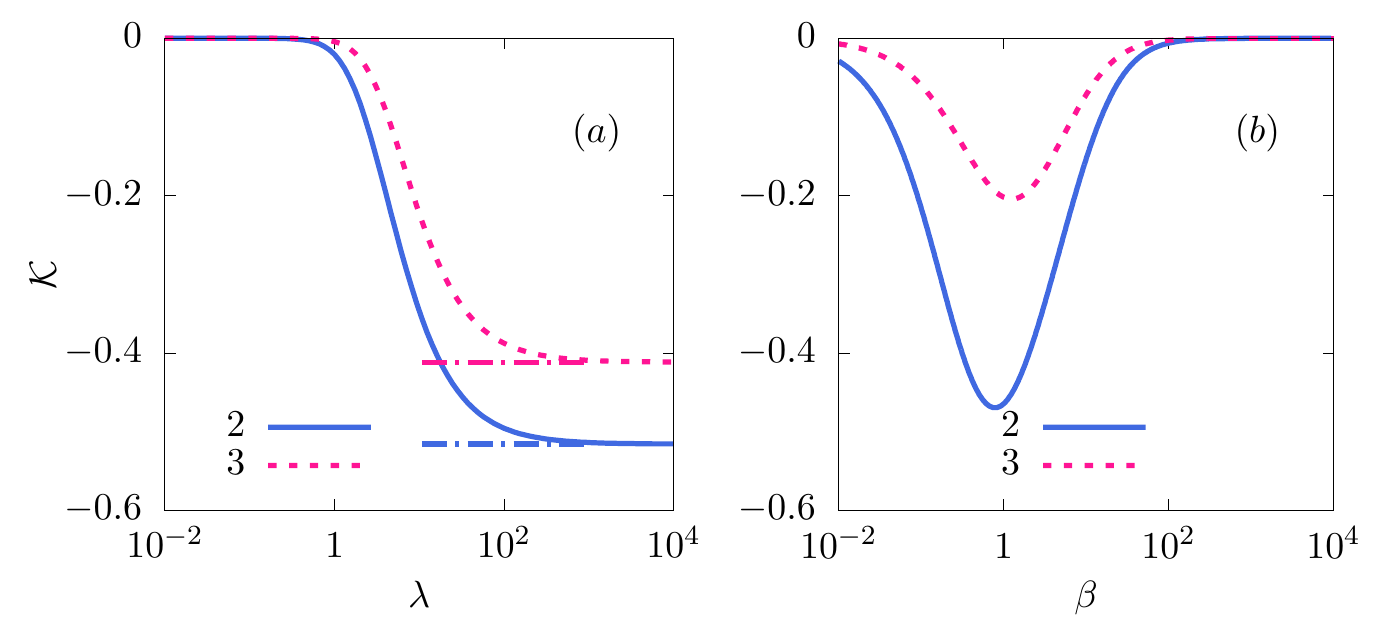} 
\caption{(color online)
The steady state kurtosis ${\cal K}$ as a function of activity $\l$~($a$), and the strength of the potential trap $\be$~($b$). In ($a$) the value of $\be=30$, and in ($b$) the value of $\l=5$ are kept fixed. 
In each graph the two lines denote the variation of ${\cal K}$ in dimensions $d=2,\,3$. The dash-dotted lines in ($a$) denote the asymptotic values of kurtosis ${\cal K} \to \f{-2 \l^4}{d^2 (d+2) \be}$ in the limit of large activity $\l \to \infty$.
}
\label{fig:kurt2}
\end{figure}
%
The displacement vector of a passive Brownian particle reaches an equilibrium Boltzmann distribution $P(\rv) \sim \exp(-U/k_B T)$ 
 in the presence of a trapping potential $U(\rv)$ and ambient temperature $T$ with $\kb$ denoting the Boltzmann constant. Thus in a Harmonic trap $U(\rv) = \hf k \rv^2$ the displacement vector would obey the Gaussian distribution
 $P(\rv)  = \left(\f{k}{2 \pi \kb T}\right)^{d/2}\exp(-k \rv^2/2 k_B T)$. 
Such a Gaussian process with $\la \rv \ra = 0$ obeys the relation $\la \rv^4 \ra = (1+2/d) \la \rv^2 \ra^2 $. 
Using the expression for $\la \rv^2 \ra$ obtained for ABP in the harmonic trap, in the definition  
\bea
\mu_4 := \left(1+ \f{2}{d} \right) \la \rv^2 \ra^2
\eea
one can define the kurtosis 
\bea
{\cal K} = \f{\la \rv^4 \ra}{\mu_4} - 1, 
\eea
which measures the deviation from the Gaussian process. 
Clearly, for a Gaussian process ${\cal K}=0$. 
In the steady state, the kurtosis of the ABP is given by 
\bea
\fl
{\cal K} = \frac{2 \mu k v_0^4 \left[\,(1-4 d) D_r -3 \mu k \right]} {(d+2)    (d D_r+\mu k)    \left[ (d-1) D_r+3 \mu k \right]} 
\times \f{1}{    \left[d^2 D D_r+d D (\mu k-D_r)+v_0^2\right]^2  } .   
\label{stdiff}
\eea
Using the dimensionless activity $\l$ and trap- stiffness $\be$, the expression can be written as 
\bea
{\cal K} = \frac{ - 2 \be \l^4 ( 4 d  +3 \be -1)}{ (d+2) (d +\be) (d-1+3 \be) \left(d^2 +d  (\be-1)+\l^2\right)^2}.
\label{eq:kurt}
\eea
In the limit of vanishing activity $\l \to 0$, the ABP behaves as a particle diffusing in the harmonic trap following the Gaussian distribution of displacement. In this limit, the kurtosis vanishes as ${\cal K}  \sim \l^4$. As $\l \to \infty$ the kurtosis saturates to 
\bea
{\cal K} = -\f{2 \be (4d+3\be -1)}{(d+2)(d+\be)(d-1+3\be)}.
\label{eq:lminf}
\eea 
Thus with increasing $\l$, the kurtosis decreases to saturate at large $\l$~(Fig.~\ref{fig:kurt2}($a$)). This shows a {\em passive} to {\em active} crossover as a function of activity. On the other hand, with the change in trap- stiffness $\be$, the kurtosis vanishes in both the limits of $\be \to 0$ and $\be \to \infty$, the two passive limits, and reaches a negative minimum for intermediate $\be$ values~(Fig.~\ref{fig:kurt2}($b$)\,). This shows the re-entrant crossover from passive to active to passive behavior with increasing trap stiffness $\be$.  
In the limit of $\be \to 0$ the kurtosis vanishes as ${\cal K} \sim -\be$. In the other limit of  $\be \to \infty$, it vanishes as 
{
\bea
{\cal K} \approx \f{-2 \l^4}{d^2 (d+2) \be^2}.
\label{Kbeinf}
\eea  }
 
\begin{figure}[!t]
\centering
\includegraphics[width=8cm]{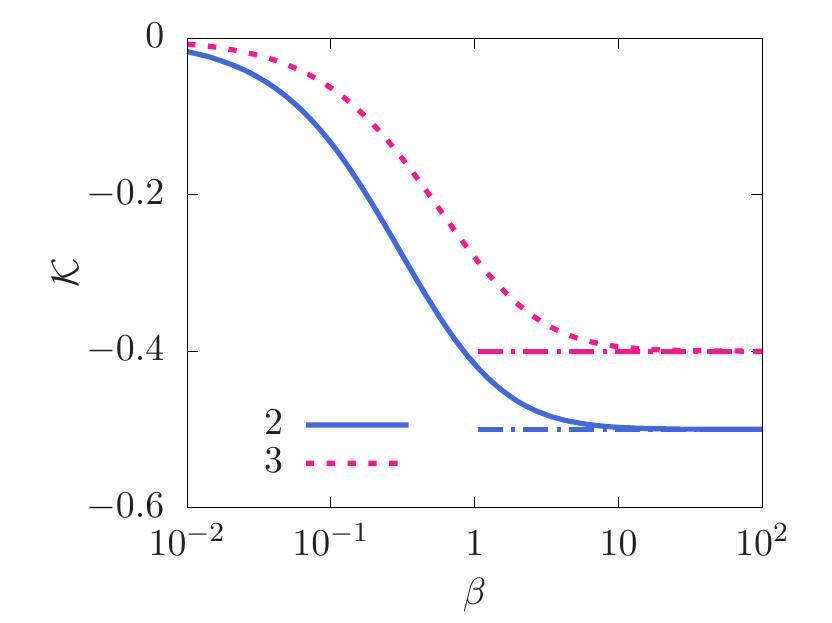}
\caption{(color online) The variation of kurtosis ${\cal K}$ with trap stiffness $\be$, in the absence of translational diffusion ($D=0$). The two graphs are for $d=2,\,3$, and the dash-dotted lines denote the asymptotic values ${\cal K} = -2/(d+2)$. { We see the monotonic dependence on $\beta$, in contrast to the non-monotonic form seen for $D \neq 0$ in Fig.~(\ref{fig:kurt2}b).} 
}
\label{fig:kurt3}
\end{figure}

 {\bf Kurtosis in the absence of translational diffusion}: 
{
The relation Eq.~(\ref{stdiff}) simplifies to the following $\l$-independent form in the limit of $D=0$, 
\bea
 {\cal K} = \frac{ - 2 \be  ( 4 d  +3 \be -1)}{ (d+2) (d +\be) (d-1+3 \be)}. 
\eea
which is the same as Eq.~(\ref{eq:lminf}), obtained in  the 
limit of $\l \to \infty$. This relation describes the  kurtosis in the system without translational diffusion. 
The above expression vanishes linearly ${\cal K} \sim - \be$ as $\be \to 0$. However, as $\be \to \infty$, it saturates to ${\cal K} = - 2/(d+2)$, in contrast to the vanishing of ${\cal K}$  at large $\be$ displayed by Eq.~\ref{Kbeinf} (compare Fig.\ref{fig:kurt3} with Fig.\ref{fig:kurt2}($b$)\,).
Thus, in the absence of translational diffusion $D$, we do not get the Gaussian behavior of the displacement distribution at large $\be$. This clearly shows that the {\em re-entrance} displayed by ${\cal K}$ as a function of $\be$ is possible only in the presence of translational diffusion $D \neq 0$.} 

{ However, the first {\em passive}  to {\em active} crossover is observed even in the absence of translational noise. The maximum radial extent of the trapped particle is given by  $r_{ac} = v_0/\mu k$. This relation is obtained by balancing the active force with the trap force.
A comparison of $r_{ac}$ with the persistence length of activity $\ell_p = v_0 \t_c = v_0/(d-1)D_r$ can be used to understand this crossover. For a shallow trapping potential, $\ell_p \ll r_{ac}$, the particle can undergo a large number of reorientations in the trap region, leading to effectively a simple diffusion in the trap and the Gaussian distribution of particle position as in equilibrium. In the other limit of stiff trapping potential, $r_{ac} \ll \ell_p$, the active particle gets localized to a separation $r_{ac}$ away from the center of the trap, leading to a strongly non-Gaussian position distribution. }

\begin{figure}[!t]
\centering
\includegraphics[width=12cm]{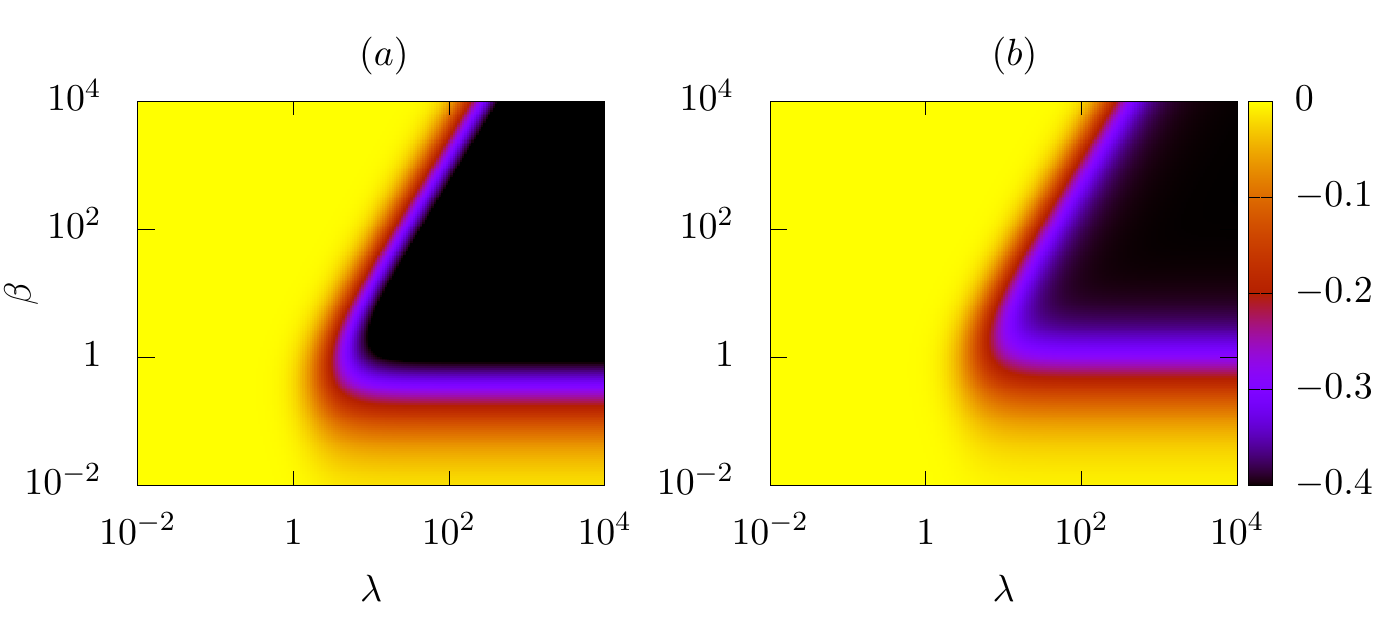}
\caption{(color online)
Heat maps of the kurtosis at steady state ${\cal K}$ as a function of the dimensionless activity $\l$ 
and potential strength $\be$
in  $d=2$~($a$) and $d=3$~($b$) dimensions. The color-box shows colors corresponding to ${\cal K}$ values shown.  
The light (yellow) and dark (black) regions denote the {\em passive} and {\em active} state, respectively.}
\label{phase_dia}
\end{figure}

 {\bf Phase diagram}: 
We show the heat-map of the kurtosis ${\cal K}$ as a function of trapping strength $\be$ and activity $\l$ in Fig.~\ref{phase_dia}, corresponding to both $d=2,\,3$ dimensions. These phase diagrams show the passive to active crossovers in the presence of finite translational diffusion.  The yellow portions of the phase space (${\cal K} \approx 0$) correspond to the {\em passive} equilibrium-like Gaussian distribution of the particle position. Whereas, the dark regions denote the largest amplitude of ${\cal K}$, capturing strong deviation from the Gaussian nature, identifying the {\em active} phase. The plot clearly shows a re-entrant transition from {\em passive}~(Gaussian) to {\em active}~(non-Gaussian) to {\em passive}~(Gaussian) behavior with increasing trap stiffness $\be$, e.g., in the region $10 \lesssim \l \lesssim 100$.

A similar phase diagram in $d=2$ was obtained earlier by directly following the nature of the probability distribution of ABP-position in a harmonic trap~\cite{Malakar2020}. Our analytic calculation of the kurtosis admits a mathematically unique description of such phase diagrams over any range of $\l$ and $\be$ values. Further, our method permits  calculation of this phase diagram in arbitrary dimensions, e.g., the {\em active}-{\em passive} transition in $d=3$ is shown in Fig.~\ref{phase_dia}($b$). 

The re-entrant transition in Fig.~\ref{phase_dia} can be understood qualitatively using the following heuristic argument~\cite{Malakar2020}. Ignoring the translational diffusion, the maximum radial extent of the trapped particle is given by  $r_{ac} = v_0/\mu k$. On the other hand, the  translational noise gives a broadened distribution around the origin, with an  associated equilibrium length scale for the spread, $\ell_{eq} = \sqrt{D/\mu k}$. The off-centered peak of the ABP  thus becomes insignificant for  $r_{ac} \lesssim \ell_{eq}$, thus giving the condition $\be \gtrsim \l^2$ for a re-entrance to the {\em passive} regime. We note that this is consistent with Eq.~(\ref{Kbeinf}).

\section{Discussion}
\label{sec:summary}
In conclusion, we have demonstrated a method of performing exact analytical calculation of all time-dependent moments of dynamical variables describing ABPs in a harmonic trap, in the presence and absence of translational diffusion. 
The non-equilibrium activity in the trapping potential leads to a position distribution away from the equilibrium Gaussian profile~\cite{Takatori2016, Malakar2020}. 
In this paper, we identified such a deviation from the Gaussian distribution in terms of the kurtosis of the displacement vector. 
Our exact analytical calculation shows a re-entrant crossover  
from a {\em passive} Gaussian to an {\em active} non-Gaussian behavior in the activity-trapping potential plane. The amplitude of kurtosis grows monotonically with increasing activity to reach a saturation value. In contrast, with increasing trap stiffness, this measure shows non-monotonic variation, reaching a maximum for intermediate trapping strengths. In both the limits of vanishing and extremely stiff limits of the harmonic potential the kurtosis vanishes. Thereby, it identifies an {\em passive} to {\em active} to {\em passive} re-entrant crossover. 
Our analytic prediction of the phase diagram, and variation of kurtosis is amenable to direct experimental verification,
e.g., using a setup similar to Ref.~\cite{Takatori2016}. 
The presence of translational diffusion is crucial for the observation of the re-entrant transition. 
In its absence, the amplitude of kurtosis increases monotonically to finally saturate with the increase of trap stiffness,
 predicting a single {\em passive} to {\em active} transition. 

\section*{Acknowledgments}
D.C. thanks SERB, India for financial support through grant number MTR/2019/000750. A.D. acknowledges support of the Department of Atomic Energy, Government of India, under project no.12-R\&D-TFR-5.10-1100. This research was supported in part by the International Centre for Theoretical Sciences (ICTS) during a visit for participating in the program - 7th Indian Statistical Physics Community Meeting (Code: ICTS/ispcm2020/02).
\vskip 2cm


\appendix

\bibliographystyle{unsrt}

\end{document}